\documentclass[twocolumn]{Boil06}
\usepackage{graphicx}
\usepackage{cite}
\title{IS CHF TRIGGERED BY THE VAPOR RECOIL EFFECT?}

\author{V. S. Nikolayev$^{1,2}$\thanks{e-mail: vadim.nikolayev@cea.fr}, D. Chatain$^1$,
D. Beysens$^{1,2}$
    \affiliation{
    $^1$ESEME, Service des Basses Temp\'eratures, DRFMC/DSM/CEA-Grenoble,
    17 rue des Martyrs, 38054 Grenoble Cedex 9, France\\
    $^2$CEA-ESEME, ESPCI-PMMH, 10, rue Vauquelin, 75231 Paris
    Cedex 5, France
    }
}
\date{This paper deals with the triggering mechanism of the boiling crisis, a transition from nucleate to film
boiling. We observe the boiling crisis in pool saturated boiling experimentally at nearly critical pressure to
take advantage of the slowness of the bubble growth and of the smallness of the Critical Heat Flux (CHF) that
defines the transition point. Such experiments require the reduced gravity conditions. Close to the CHF, the
slow growth of the individual dry spots and their subsequent fusion on the transparent heater are observed
through the latter. As discussed in the paper, these observations are consistent with numerical results obtained
with the vapor recoil model of the boiling crisis.}

\begin{document}

\maketitle

\section*{INTRODUCTION}

The boiling crisis, called also the ``Departure from Nucleate Boiling" (DNB), is a transition from nucleate to
film boiling. It is highly important for industrial applications of high heat flux boiling because of the rapid
heat transfer decline at the moment of transition. A severe damage of the heater can result because of its
overheating. DNB occurs at a threshold value of the heat flux, the Critical Heat Flux (CHF). Because of its
importance, a large number of the CHF studies is being published each year since the first model suggested by
Zuber \cite{Zuber}. While Zuber's CHF expression describes relatively well a number of experimental CHF data,
the underlying physical model (vapor stems) does not correspond to most modern observations of DNB. Multiple
semi-empirical correlations were proposed since that. However, the validity range for each corresponds basically
to the range where the empirical constants were fitted. The predictive power of such correlations remains thus
very limited.

All existing modern ideas (see e.g. \cite{Buyevich} for their review) concerning the DNB triggering can be
placed into two large classes. The first considers the heater drying is initiated by the growth of a single
bubble as the primary mechanism. The second argues that the bubble crowding close to the heater leads to the
fusion of multiple small dry spots. This paper develops an approach of the first type based on the vapor recoil
concept proposed originally in \cite{EuLet99}. The numerical simulations are performed. We discuss also recent
experimental observations which corroborate this idea.

For the reasons of industrial importance, boiling is mostly studied at low pressures comparing to the critical
pressure $p_c$ (the pressure of the gas-liquid critical point) of the fluid under study, e.g. for water or freon
at atmospheric pressure. CHF is then large (of the order of several MW/m$^2$ for water) and the boiling close to
it is indeed extremely violent. However, it is well known that CHF decreases at high pressures where DNB can
thus be observed at a smaller heat flux. In addition, the thermal diffusivity is smaller in this regime, the
bubble growth slows down, and the optical distortions disappear due to the slowness of the fluid motion. From
the point of view of the modelling, this slowness permits to simplify the problem by neglecting the hydrodynamic
stresses at the bubble interface, the shape of which can then be calculated in the quasi-static approximation.

\section*{VAPOR RECOIL EFFECT ON THE BUBBLE SPREADING}

Every fluid molecule evaporated from the liquid interface causes a recoil force analogous to that created by the
gas emitted by a rocket engine. It pushes the interface towards the liquid side in the normal direction. The
vapor recoil force appears because the fluid necessarily expands while transforming from liquid to gas phase.
Obviously, the stronger the evaporation rate $\eta$ (mass per time and interface area), the larger the vapor
recoil force. The vapor recoil force per unit interface area is $P_r=\eta^2(\rho_V^{-1}-\rho_L^{-1})$, where
$\rho_L$($\rho_V$) signifies liquid (vapor) density.

\begin{figure}[htb]
\centering
\includegraphics[width=6cm]{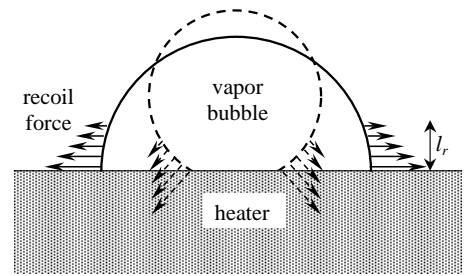}
\caption{Sketch illustrating the vapor recoil effect. The amplitude and direction of the vapor recoil force are
shown by arrows. The thickness $l_r$ of the belt on the bubble surface, in which the vapor recoil force is
important, is exaggerated with respect to the bubble size.} \label{brec}
\end{figure}
\begin{figure}[htb] \centering
\includegraphics[width=8.5cm]{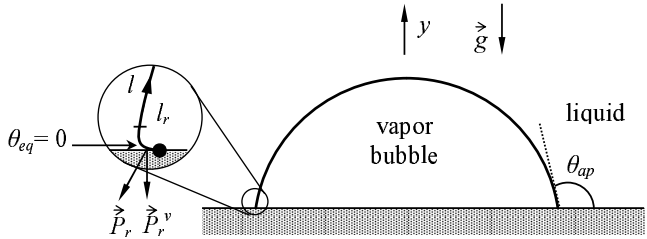}
\caption{Apparent $(\theta_{ap})$ and actual $(\theta_{eq})$ contact angles. The vicinity of the contact line
(zero for the curvilinear coordinate $l$ measured along the bubble contour) is zoomed in. The arrows show the
directions of the vapor recoil force and its vertical component responsible for an additional adhesion of the
bubble to the heater.} \label{appangle}
\end{figure}

Let us now consider a growing vapor bubble attached to the heater surface (Fig.~\ref{brec}). While the
temperature of the vapor-liquid interface is constant and equal to the saturation temperature for the given
system pressure $p$ for the pure fluid case (see the discussion in \cite{Straub,PRE01,Tadrist} and references
therein), a strong temperature gradient forms in the liquid near the heating surface. The liquid is overheated
in a thermal boundary layer, and the heat flux $q_L$ at the bubble surface is thus elevated in a ``belt" of the
bubble surface adjacent to the bubble foot. As a matter of fact, most of the evaporation into the vapor bubble
is produced in this belt, whose thickness $l_r$ is much smaller than the bubble radius. Since $\eta=q_L/H$,
where $H$ is the latent heat, the vapor recoil near the contact line is much larger than at the other part of
the bubble surface and the contact line is pulled apart from the bubble center as if the contact angle
increased. However the actual contact angle $\theta_{eq}$ depends only on the molecular forces and remains
constant. The bubble curvature should thus increase near the contact line, see Fig.~\ref{appangle}.

In other words, the vapor recoil tends to extend the dry spot under the vapor bubble so that the bubble tends to
cover the heater surface. However other forces acting on the bubble surface can counterbalance the vapor recoil.
E. g. for the usual evaporation conditions, the vapor recoil is negligibly small with respect to the surface
tension $\sigma$. Their relative magnitude is characterized by the ratio
\begin{equation}\label{N}
N=\int P_r(l)\mathrm{d}l/\sigma,
\end{equation}
where the integration is performed along the bubble contour, see Fig.~\ref{appangle} for the $l$ definition. The
spreading occurs when $N$ becomes comparable with unity which corresponds to the heat flux from the heater of
the order 1 MW/m$^2$ \cite{EuLet99} for water at high pressure, i.e. to the flux comparable to the actual CHF.
The spreading of a bubble is followed by the coalescence with its neighbors.

The effect of the vapor recoil is not limited to the bubble spreading. The vapor recoil creates an additional
bubble adhesion to the heater that prevents the bubble departure as soon as the bubble spreading begins, i.e.
when the vapor recoil becomes important \cite{MI04}. This adhesion force can be obtained by integrating the
vertical component $P_r^v$ of $\vec{P}_r$ (Fig.~\ref{appangle}) over the bubble interface. This point is
extremely important because the CHF should be determined as a threshold between the bubble departure and
spreading regimes. If the time of bubble residence on the heater is small, the bubble might simply do not have
enough time to spread.

These considerations would not allow a CHF correlation to be obtained analytically. Thus the numerical
simulations are necessary.

\section*{NUMERICAL MODELLING}

The account of the vapor recoil effect requires the simulation of the bubble growth with a truly free surface.
This complicates the simulation a lot. However, there is a one more problem associated with a singular (or
quasi-singular) behavior of the local heat fluxes in the vicinity of the contact line. Indeed, since the thermal
conductivity of the heater is usually much larger than that of the liquid, the heater is assumed isothermal at a
temperature higher than the saturation temperature $T_{sat}$ for the given system pressure. If the bubble
surface is assumed isothermal at $T_{sat}$ in agreement with the experimental and theoretical considerations
\cite{Straub,PRE01,Tadrist}, the resulting temperature ambiguity at the contact line leads to the non-integrable
divergence, $q_L(l)\sim l^{-1}$. This shows importance of the thermal properties of the heater that need to be
taken into account by solving the conjugate problem at least in liquid and solid domains.

\subsection*{Problem statement and its solution}

At high pressures we can assume that the bubble growth is slow and than the viscous stresses at the bubble
interface and the inertial expansion forces are much smaller than the surface tension. In this case the bubble
shape is independent of hydrodynamics any more and only heat conduction problem can be solved in the first
approximation to describe the bubble growth. The bubble shape is then determined out of the quasi-static
approach that consists to assume that at each time moment bubble has a shape as if it were in equilibrium for
the given instantaneous force distribution along its surface. The bubble shape comes from the value of the
bubble local curvature $K(l)$ satisfying the modified Laplace equation
\begin{equation} K\sigma=\lambda+(\rho_L-\rho_V) g
y+P_r,\label{surf}
\end{equation}
where $g$ is the gravity acceleration directed as shown in Fig.~\ref{appangle}, $y$ is the ordinate, and
$\lambda$ is the constant along the bubble surface pressure difference between the pressures inside and outside
of the bubble. The imposed contact angle $\theta_{eq}$ gives a required boundary condition for the shape
determination. $\lambda$ plays a role of the Lagrange multiplier determined knowing the bubble volume $V$ which
is defined by the total amount of the latent heat consumed by the growing bubble
\begin{equation}
H\rho_V{{\rm d}V\over{\rm d}t}=\int q_L(l) \;{\rm d}l. \label{eqnV}
\end{equation}
The 2D case is assumed hereafter. The only yet undefined quantity in the described problem is $q_L(l)$ is found
from the coupled at the heater-liquid interface heat conduction problems in the heater and the liquid (that with
moving boundary), both semi-infinite. The vapor is assumed non-conductive and zero heat flux condition is
imposed at the dry spot, i.e. at the heater-vapor contact area. The heat is assumed to be generated
homogeneously (as by electric current) in the heater volume. The time variation of this volume source is chosen
in such a way that the heat flux $q_0$ at the heater-liquid boundary far from the bubble remains constant in
time. The $q_0$ value will be used as the main control parameter. The constant temperature $T_{sat}$ is given as
the boundary condition at the moving bubble interface and also as the initial condition. A small vapor bubble of
the radius $R_0$ is assumed to exist prior to the calculation beginning.

This free boundary problem is solved with the Boundary Element Method (BEM). More details on the solution and on
the method can be found in \cite{IJHMT01}.

The evaporation heat flux $q_L(l)$ remains singular at small $l$ in this case, however its divergence is
integrable since the exponent $\alpha$ of $q_L(l)\sim l^\alpha$ always falls between $-0.5$ and $-0.8$. At the
same time $P_r\propto q_l^2$ is non-integrable which leads to a difficulty of the shape determination
\footnote{This is easy to understand by recalling the strong dependence of the bubble shape on $N$ given by
(\ref{N}) that involves the $P_r$ integral}. To overcome this problem, a cut-off in the flux is needed. From the
physics point of view, $q_l$ is bounded by $q_{max}$ \cite{IJHMT01} due to the limitation on the heat transfer
imposed by the maximum molecular speed that introduces a thermal resistance at the bubble interface. Since this
cut-off is much larger than the values encountered at the node points used in the numerical calculation, no
modification of the isothermal boundary condition is necessary. The same sensitivity of the shape on the $P_r$
integral implies a requirement of the high accuracy of the $\alpha$ determination and thus very fine meshing (of
the order of $10^{-4}$ of the bubble shape) in the contact line vicinity which influences the calculation time.
This reason determined our choice of BEM where the node points' number is much less than in any other numerical
method: only boundaries between the liquid, vapor and solid domains need to be meshed, not the domains
themselves. As a matter of fact, even such fine meshing is not sufficient. To achieve the satisfactory accurate
$P_r$ integral calculation, the $l^\alpha$ extrapolation of $q_l$ must be used till $q_{max}$ is reached. The
corresponding $P_r$ contribution is integrated analytically and added to the numerically integrated part.

The numerical algorithm has been improved with respect to that used in \cite{IJHMT01} to get rid of the
numerical instabilities which gave rise to temporal oscillations visible e.g. in figure 6 of \cite{IJHMT01}, to
be compared with Fig.~\ref{Grav} below. In particular, a new algorithm \cite{Polev} was used to solve Eq.
\ref{surf}.

\subsection*{Results}

The bubble spreading has already been shown in \cite{IJHMT01} where $\theta_{eq}=0$ was assumed and no forces
tending to remove the bubble from the heater were taken into account. In this paper we introduce the gravity
that appears in Eq. \ref{surf}. To understand the gravity influence, let us first assume that $P_r=0$ in Eq.
\ref{surf}. The latter has then no solution (i.e. the equilibrium bubble shape) for large $V$ when the contact
angle is fixed. A solution of such a truncated problem exist only when $V$ is smaller than some maximum value
$V_{max}=V_{max}(\theta_{eq})$ \cite{Troyes}, when the capillary adhesion is larger than the Archimedes force.
At small $\theta_{eq}$, when $V$ approaches $V_{max}$ during the bubble growth, the dry area size tends to zero
and at $V_{max}$ no adhesion to the heater exists any more which means that the bubble departs as a whole.

\begin{figure}[htb] \centering
\includegraphics[width=8cm]{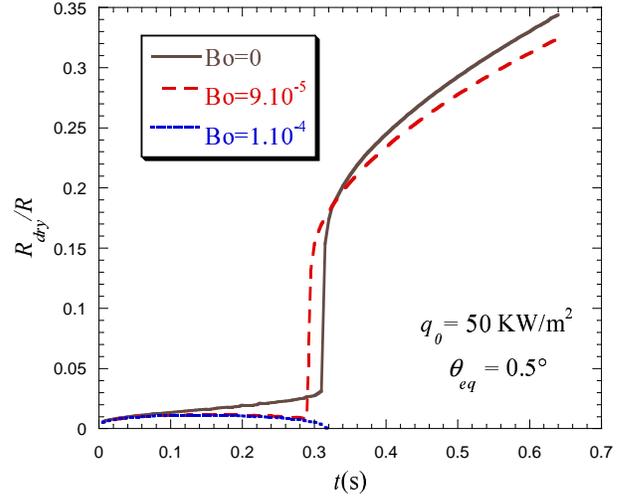}
\caption{The dry spot evolution with respect to the bubble radius $R$ for fixed heat flux and different gravity
levels calculated for the same parameters as in \cite{IJHMT01} (water at 10 MPa).} \label{Grav}
\end{figure}

The account of $P_r$ adds another option due to the $P_r^v$ influence discussed above (Fig.~\ref{Grav}).
Either the dry spot collapses and the bubble departs as in the $P_r=0$ case, or its spreading begins. The vapor
recoil adhesion then exceeds at once the Archimedes force and keeps increasing with time \cite{MI04} so that the
bubble continues to spread without departing from the heater. This second scenario corresponds to that of the
boiling crisis. The lowest value of $q_0$ at which it occurs (i.e. the transition between the two scenarios)
gives the model CHF.

Fig.~\ref{Grav} shows the behavior of the dry spot at a fixed heat flux and a different gravity levels given by
the Bond number Bo$=(\rho_L-\rho_V)gR_0^2/\sigma$. The spreading occurs at Bo$=9\cdot 10^{-5}$ (which
corresponds to about $0.1g$) while at Bo$=10^{-4}$ the bubble departs. It means that the CHF is 50 KW/m$^2$ for
the chosen system parameters.

Note that the insensitivity of the CHF to gravity \cite{Straub} at low pressures can be explained in the
framework of the present model by the influence of the inertial hydrodynamic forces that cause the bubble
departure even in the absence of gravity. According to the present model, when the gravity contribution is
smaller than that of the inertia, the CHF should be rather defined by the latter than by the former. We will see
in the next section that for high pressures, when the bubble growth is slow and the inertial forces are
negligible, the CHF is indeed sensitive to the gravity.

\section*{EXPERIMENTS AT EXTREMELY HIGH PRESSURE}

To overcome the experimental difficulties encountered during the observations of DNB at low pressures, we carry
out our experiments at very high pressures, in the vicinity of the critical point defined by the critical
pressure $p_c$ and temperature $T_c$. We take advantage of the so-called ``critical slowing down" observed near
the critical point. In fact, due to the smallness of the thermal diffusivity, the growing process of a single
vapor bubble could be observed during minutes thus allowing for a very detailed analysis. The CHF value is also
vanishing at the critical point \cite{Tong} so that DNB can be examined at a small heat flux that does not
necessarily induce a strong fluid motion and high temperature gradients which hinder the optical observations.
However, near-critical bubble growth experiments have an important drawback. Since the surface tension becomes
very low near the critical point, gravity completely flattens the liquid interface. Reduced gravity conditions
are thus necessary to preserve the bubble shape.

In our already performed experiments, the cells are closed and only pure fluid is present in them so that its
total mass and volume remains constant. Unlike the conventional boiling experiments, the gas bubble is not
nucleated but exists already before the cell heating begins. The bubble growth is then observed.

A particularity of a near-critical fluid consists in the symmetry of its co-existence curve (temperature
dependence of $\rho_L$ and $\rho_V$) with respect to the critical density $\rho_c$: $(\rho_L+\rho_V)/2=\rho_c$.
When the average density is equal to $\rho_c$, the vapor volume remains nearly constant and equal to one half of
the cell volume throughout the heating. This makes the optical observations even more convenient. The vapor mass
however increases (and the liquid mass decreases) during the heating because of the density change.

Two kinds of experiments were carried out up to now. The first studies \cite{PRE01} used SF$_6$ fluid on board
of the Mir space station in the ALICE-2 apparatus designed by the French CNES agency. The choice of SF$_6$ is
made for practical reasons: the critical point of this fluid is $T_c=45.6^\circ$C, $p_c=3.8$~MPa and requires
much less severe conditions for the experiment than for example water ($374^\circ$C, 22~MPa). The sequential
photos of the growing vapor bubble showed its spreading over the heater. The increase of the apparent contact
angle was clearly seen while the actual contact angle was exactly zero (which is a common feature of
near-critical fluids).

\begin{figure}[htb] \centering
\includegraphics[width=8cm]{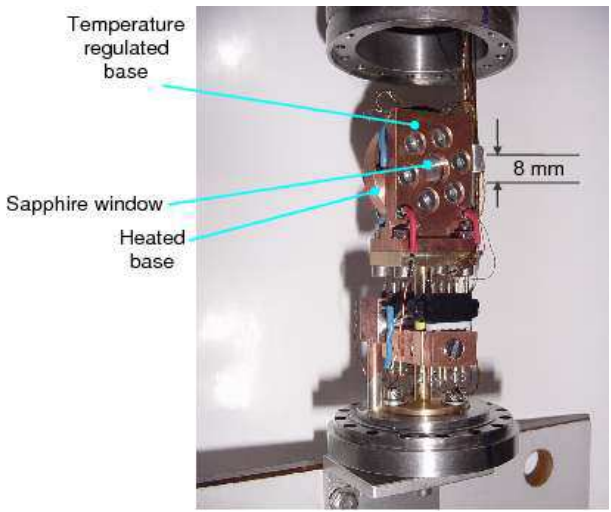}
\caption{H$_2$ cell situated inside an "anti-cryostat" under vacuum which is in its turn situated inside another
cryostat filled by liquid He to cool the system.} \label{Cell}
\end{figure}

However, the cells in ALICE-2 were not suitable to control the heat supply or to measure it to obtain the
quantitative data. This limitation has been overcome in the experimental H$_2$ setup that makes use of the
magnetic levitation facility \cite{Mag} at CEA-Grenoble. All further description will concern this data. A
cylindrical cell (Fig.~\ref{Cell})
of 8~mm diameter and 5~mm height is filled with H$_2$ at critical density. The sapphire windows (cylinder bases)
are good heat conductors in the cryogenic temperature range ($T_c=33$K for H$_2$). The lateral cell wall is made
of stainless steel which is on the contrary the thermal insulator, its heat conductivity being about 1000 times
less than that of the sapphire. The copper rings serve to transfer the heat to the windows and to keep the parts
of the cell together. The fluid is heated by one of the windows while the temperature of the other is maintained
by the temperature regulation system that also permits us to measure the heat flux removed from the cell. The
cell can be observed optically through the transparent heater with a light source and a camera. Both are
situated outside the cryostat. The optical links are made with the help of light guides.

The gravity is compensated by magnetic forces within 2\% in the cell volume. The position of the cell with
respect to the magnetic field is chosen so that the residual force positions the bubble against the heating
window. In the window plane, the residual force provides the effective gravity directed from the periphery of
the window to its center. This force field leads to a curious phase distribution close to the critical point
when the surface tension decreases. Most of the more dense liquid phase then masses in the center while the
wetting film remains at the cell walls and the windows. As a consequence, the bubble forms a torus (we call this
topology annular because of the observed bubble contour). Farther from the critical point, the bubble remains
circular as usually.

\begin{figure}[htb]
\centering
\includegraphics[width=8cm]{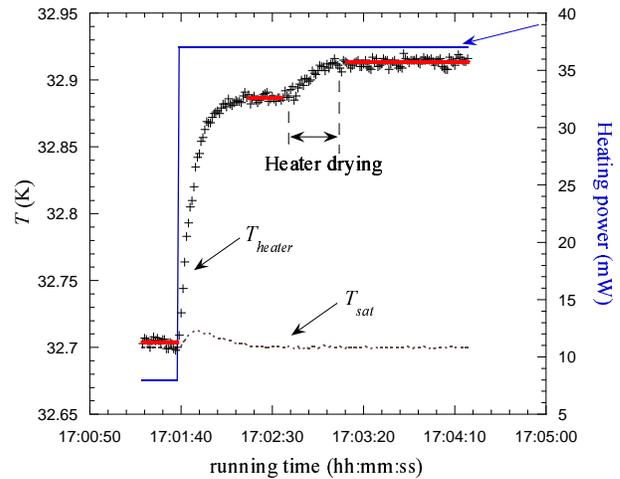}
\caption{Heater temperature evolution at CHF (crosses). The thick horizontal lines correspond to the averaged
values at tree stationary states: at equilibrium, just before and after the DNB. The dotted line reflects the
system pressure evolution given by $T_{sat}$.} \label{T_heater}
\end{figure}

\begin{figure}[htb]
\centering
\includegraphics[width=8cm]{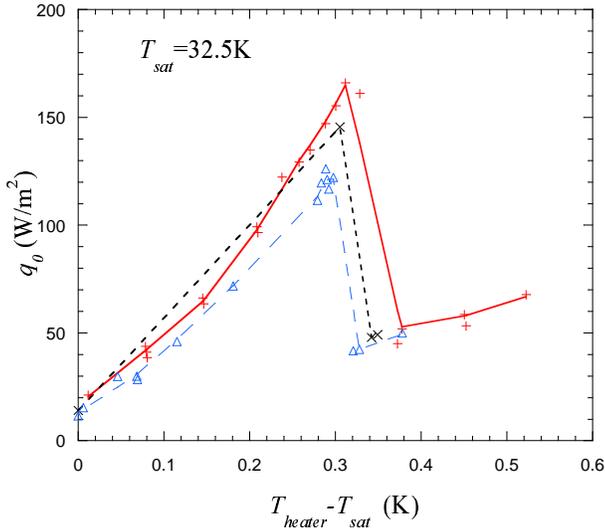}
\caption{Boiling curves for the same pressure but different effective gravity levels.} \label{boilcurves}
\end{figure}

The heater temperature evolution at DNB is shown in Fig.~\ref{T_heater}.
Before the observation, the heating power is adjusted to compensate for the heat losses and maintained
sufficiently long to achieve the thermal equilibration in the isothermal state. When the heat injection starts,
small bubbles are nucleated and depart from the heater under the action of residual gravity before coalescing
with the large bubble. When the flux approaches the CHF, the dry spot under bubbles nucleates and attains a
finite value. At CHF the dry spots under virtually each of the bubbles begin to grow. The larger the dry spot,
the faster the growth. At the last stage, the dry spots coalesce and the dryout occurs rapidly. At CHF the
complete heater drying (DNB) occurs after a waiting time during which the heater temperature exhibits
fluctuations. DNB occurs after the largest of them. The heater is completely dried out and a new stationary
conductive state (no convection) is quickly established. In this state, the main part of the heat passes from
one base to another by the cell walls. This heat transfer was accounted for by measuring the heat transfer in
the empty cell and subtracting the wall flux from the total measured heat flux. The resulting boiling curves are
presented in Fig.~\ref{boilcurves}.
It shows the dependence of the CHF on very slight changes of the bubble position with respect to the heater.
These changes can be induced by slightly displacing the cell with respect to the field so that the wetting film
existing at equilibrium between the bubble and the heater, changes its thickness. This sensitivity manifests
itself even stronger when the geometry of the large bubble changes from circular to annular. During this change
the wetting layer becomes much thicker and occupies smaller portion of the window. The thickness can be judged
from the size of the small bubbles that grow inside the layer. After the geometry change occurs, the CHF varies
strongly.

This change of bubble position is analogous to the change in the force balance acting on the drop in the usual
boiling set-up where there is always a dry spot, at least in a nucleation site area. This dependence can be
summarized as follows: the smaller adhesion, the larger is the CHF. This observation is coherent with our vapor
recoil model.

\begin{figure}[htb]
\centering
\includegraphics[width=8cm]{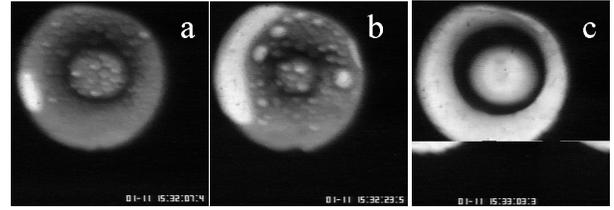}
\caption{Gas spreading at 32.95K (at annular geometry, the large bubble has a toroidal shape as described in the
text) as visualized through the transparent heater in magnetic levitation experiment. (a) Beginning of the dry
spot growth. (b) Bubble partially spread. (c) Complete drying of the heater. Nucleated small bubbles are visible
in (a-b).} \label{dry_spot}
\end{figure}

Optical observations of the dry spot growth are very interesting. They are especially informative for the
annular bubble geometry because the small bubbles are nucleated in the central region of the window, their
dynamics can thus be compared to that of the large bubble. For the circular bubble case (farther from the
critical point), the small bubbles nucleate and grow only at the periphery of the cell where the liquid layer is
thicker. Far from the CHF, the small bubbles first depart from the heater and then coalesce with the large
bubble. Their immediate absorption by the large bubble follows. Closer to the CHF, the dry spot under the large
bubble forms. The small bubbles also develop a dry spot comparable with their size. Once the dry spot begins to
grow, the bubble would not depart as quickly. It would rather continue to grow (in the agreement with our model)
until it touches the large bubble and quickly absorbed by it. The growing bubbles in the center slide towards
the periphery of the window (probably, because of the larger radial gravity component in the center). The large
bubble exhibits the largest rate of the dry spot growth at CHF (Fig.~\ref{dry_spot}).

First, the single dry spot under the large bubble spreads (Fig.~\ref{dry_spot}a), then it coalesces with small
bubbles so that their dry spots fuses (Fig.~\ref{dry_spot}b). Ultimately, the whole heater surface dries out
during rapid coalescence motions and the fluid becomes very turbid. After the slow relaxation to the transparent
state, a stationary conductive state in which the vapor phase covers the heater and the interface forms the
hat-like shape (Fig.~\ref{dry_spot}c) where the fluid layer in the center of the window is thicker than at the
periphery.

The similar mechanism of the dry spot spreading works also relatively far from the critical point where the
bubble is circular. Indeed, the experiments were performed for up to 3\% deviation from $p_c$ (which is well
beyond the critical region where the anomalous fluctuations define completely the behavior of the fluid).

\section*{CONCLUSIONS}

Basing on the vapor recoil model, the individual bubble spreading is obtained in the simulations. To capture the
vapor recoil effect, an extremely delicate analysis of the heat flux maximum in the vicinity of the critical
point need to be performed. A threshold exists between the spreading and the bubble departure regimes. This
threshold heat flux can be associated with the CHF. This kind of simulation allows to predict the dependence of
the threshold on various parameters of the system like contact angle, thermal properties of the heater and the
liquid, etc.

The experiments at nearly critical pressure can give very detailed information about the DNB because of the
small CHF value and the slowness of the bubble growth. The reduced gravity conditions are however required to
observe the bubbles. During the DNB, we observed clearly the growth of individual dry spots under the bubbles
prior to their coalescence. This allows us to suggest spreading of individual bubbles as the triggering
mechanism for the DNB. This picture agrees with the vapor recoil model suggested earlier.

%%%%%%%%%%%%%%%%%%%%%%%%%%%%%%%%%%%%%%%%%%%%%%%%%%%%%%%%%
\begin{acknowledgment}
We acknowledge the partial financial support from CNES. We are grateful  to V. K. Polevikov for the helpful
correspondence.
\end{acknowledgment}

%%%%%%%%%%%%%%%%%%%%%%%%%%%%%%%%%%%%%%%%%%%%%%%%%%%%%%%%%
\begin{nomenclature}
\entry{Bo}{Bond number}%
\entry{$g$}{gravity acceleration [m$^2$/s]}%
\entry{$H$}{latent heat [J/kg]}%
\entry{$K$}{curvature [m$^{-1}$]}%
\entry{$l$}{coordinate varying along the bubble contour}%
\entry{$N$}{vapor recoil strength}%
\entry{$p$}{pressure [N/m$^2$]}%
\entry{$P_r$}{vapor recoil pressure [N/m$^2$]}%
\entry{$q$}{heat flux [W/m$^2$]}%
\entry{$R$}{bubble radius [m]}%
\entry{$t$}{time [s]}%
\entry{$T$}{temperature [K]}%
\entry{$V$}{2D-bubble volume [m$^2$]}%
\entry{$y$}{ordinate}%
\entry{$\alpha$}{$q_L$ exponent}%
\entry{$\eta$}{rate of evaporation [kg/(s$\cdot$m$^2$)]} %
\entry{$\theta$}{liquid contact angle}%
\entry{$\lambda$}{vapor/liquid pressure difference [N/m$^2$]}%
\entry{$\rho$}{mass density [kg/m$^3$]}%
\entry{$\sigma$}{surface tension [N/m]}%
\entry{sat}{saturation}%
\entry{c}{critical}%
\entry{v}{vertical}%
\entry{eq}{equilibrium}%
\entry{ap}{apparent}%
\entry{0}{initial}%
\entry{max}{maximum}%
\entry{V}{vapor}%
\entry{L}{liquid}%
\end{nomenclature}

%%%%%%%%%%%%%%%%%%%%%%%%%%%%%%%%%%%%%%%%%%%%%%%%%%%%%%%%%

\endmytext %this - unbelievably - balances columns in last page

\end{document}